\documentclass[11pt,a4paper]{article}
\usepackage{moriond,epsfig}

\begin{document}
\vspace*{4cm}
\title{The study of charmless hadronic $B_s$ decays}

\author{ Cai-Dian L\"u}
\address{Institute of High Energy
Physics, CAS, P.O.Box 918(4), Beijing 100049, China}

\maketitle\abstracts{ The  perturbative QCD approach has achieved
great success in the study of hadronic B decays. Utilizing the
constrained parameters in these well measured decay channels, we
study most of the possible charmless $B_s \to PP$, $PV$ and $VV$
decay channels in the perturbative QCD approach. In addition to the
branching ratios and CP asymmetries, we also give predictions to the
polarization fractions of the vector meson final states. The size of
SU(3) breaking effect is also discussed. All of these predictions
can be
tested by the future   LHCb experiment.} %
\noindent {\small¥{\it Keywords}: $B_s$ decays, factorization, pQCD

\section{Introduction}

There is a continuous progress in the study of hadronic B decays
since the so called naive factorization approach.\cite{bsw,9804363}
In recent years, the QCD factorization approach (QCDF) \cite{qcdf}
and perturbative QCD factorization (pQCD) approach \cite{pqcd}
together with the soft-collinear effective theory \cite{scet} solved
a lot of problems in the non-leptonic decays. Although most of the
branching ratios measured by the B factory experiments can be
explained by any of the theories, the direct CP asymmetries measured
by the experiments are ever predicted with the right sign only by
the pQCD approach.\cite{direct} The LHCb experiment will soon run in
the end of 2007. With a very large luminosity,  it will accumulate a
lot of $B_s$ events. The progress in both theory and experiment
encourages us to apply the pQCD approach to the charmless $B_s$
decays in this work.\cite{PQCDBs}

In the hadronic $B(B_s)$ decays, there are various energy scales
involved. The factorization theorem allows us to calculate them
separately. First, the physics from the electroweak scale down to b
quark mass scale is described by the renormalization group running
of the Wilson coefficients of effective four quark operators.
Secondly, the hard scale from b quark mass scale to the
factorization scale $\sqrt{\Lambda m_B}$ are calculated by the hard
part calculation in the perturbative QCD approach.\cite{li2003} When
doing the integration of the momentum fraction $x$ of the light
quark, end point singularity will appear in the collinear
factorization (QCDF and SCET) which breaks down the factorization
theorem. In the pQCD approach, we do not neglect the transverse
momentum $k_T$ of the light quarks in meson. Therefore the endpoint
singularity disappears. The inclusion of transverse momentum will
also give large double logarithms ln$^2k_T$ and ln$^2x$ in the hard
part calculations. Using the renormalization group equation, we can
resum them for all loops to the leading order resulting Sudakov
factors. The Sudakov factors suppress the endpoint contributions to
make the calculation consistent.\cite{pqcd}

 The physics below the
factorization scale is non-perturbative in nature, which is
described by the hadronic wave functions of mesons. They are not
perturbatively calculable, but universal for all the decay
processes. Since many of the hadronic
 and semi-leptonic $B$ decays have been measured well in the two B
 factory experiments, the light wave functions are strictly
 constrained.
  Therefore, it is useful to use the same wave functions in our  $B_s$
decays determined from the hadronic $B$ decays. The uncertainty of
the hadronic wave functions will come mainly from the SU(3) breaking
effect between the $B_s$ wave function and $B$ wave
function.\cite{PQCDBs} In practice, we use a little larger
$\omega_b$ parameter for the $B_s$ meson than the $B_d$ meson, which
characterize the fact that the light $s$ quark in $B_s$ meson
carries a littler larger momentum fraction that the $d$ quark in the
$B_d$ meson.

 \section{Results and Discussion}

For $B_s$ meson decays with two light  mesons in the final states,
the light mesons obtain large momentum of 2.6GeV in the $B_s$ meson
rest frame. All the quarks inside the light mesons are therefore
energetic and collinear like. Since the heavy b quark in $B_s$ meson
carry most of the energy of $B_s$ meson, the light $s$ quark in
$B_s$ meson is soft. In the usual emission diagram of $B_s$ decays,
this quark goes to the final state meson  without electroweak
interaction with other quarks, which is called a spectator quark.
Therefore there must be a connecting hard gluon to make it from soft
like to collinear like.
   The hard part of the interaction
becomes six quark operator rather than four. The soft dynamics here
is included in the meson wave functions. The decay amplitude is
infrared safe and can be factorized as the following formalism:
\begin{equation}
C(t) \times H(t) \times \Phi (x) \times \exp\left[ -s(P,b) -2 \int
_{1/b}^t \frac{ d \bar\mu}{\bar \mu} \gamma_q (\alpha_s (\bar \mu))
\right], \label{eq:factorization_formula}
\end{equation}
where $C(t)$ are the corresponding Wilson coefficients of four quark
operators, $\Phi (x)$ are the meson wave functions and the variable
$t$ denotes the largest energy scale of hard process $H$, which is
the typical energy scale in PQCD approach and the Wilson
coefficients are evolved to this scale.  The exponential of $S$
function is the so-called Sudakov form factor resulting from the
resummation of double logarithms occurred in the QCD loop
corrections, which can suppress the contribution from the
non-perturbative region.  Since logarithm corrections have been
summed by renormalization group equations, the
 above factorization formula does not depend on the renormalization
scale $\mu$ explicitly.

 The numerical results of the $B_s$ decays
branching ratios and CP asymmetry parameters  are displayed in
ref.\cite{PQCDBs} In all the decay channels for charmless $B_s$
decays, only several are measured by the CDF
collaboration.\cite{CDFBsKKrecent}  We show those channels together
with results by QCDF \cite{QCDFBs1} and SCET approaches
\cite{SCETBs} in table~\ref{tab:exp}. From those comparison, we
notice that the measured branching ratios are still consistent with
the theoretical calculations. Like the case in $B$ decays, the
calculated branching ratios from the three kinds of methods overlap
with each other, considering the still large  theoretical
uncertainties. A global fit is useful when we have enough measured
channels.

In table~\ref{tab:exp}, the only measured CP asymmetry in $B_s \to
K^- \pi^+$ decay prefer our pQCD approach rather than QCDF approach.
This is similar with the situation in $B$ decays. The direct CP
asymmetry is proportional to the sine of the strong phase difference
of two  decay topologies.\cite{direct} The strong phase in our pQCD
approach is mainly from the chirally enhanced space-like penguin
diagram, while in the QCDF approach, the strong phase mainly comes
from the virtual charm quark loop diagrams. The different origin of
strong phases gives different sign to the direct CP asymmetry imply
a fact that the dominant strong phase in the charmless decays should
come from the annihilation diagrams.  It should be noted that the
SCET approach can not predict the direct CP asymmetry of $B$ decays
directly, since they need more experimental measurements as input.
However, it also gives the right CP asymmetry for $B_s$ decay if
with the input of experimental CP asymmetries of $B$ decays, which
means good SU(3) symmetry here.

\begin{table}[t]
\caption{The branching ratios and CP asymmetry  calculated in PQCD
approach, QCDF
 and SCET approaches together with Experimental Data.\label{tab:exp}} \vspace{0.1cm}
\begin{center}
\begin{tabular}{|c|c|c|c|l|}
\hline & SCET  &     QCDF   &   PQCD    &   EXP
\\ \hline
$B(B_s\to K^- \pi^+) (  10^{-6})$ &  $4.9\pm 1.8$  & $  10\pm6 $& $
11 \pm 6$
& $5.0\pm 1.3$\\
$B( B_s\to K^- K^+)(  10^{-6})$ &$18\pm 7$  &    $ 23\pm 27$ & $17
\pm9$&
$24 \pm 5 $ \\
$B(B_s\to \phi \phi)(  10^{-6})$ & & $22 \pm30$&$ 33 \pm13$&$ 14 \pm
8$
 \\
\hline $A_{CP}(B_s\to K^- \pi^+)$ (\%) &$20 \pm 26$ & $     -6.7
\pm16$& $ 30 \pm 6$&$    39 \pm15 \pm 8$
\\ \hline
\end{tabular}
\end{center}
\end{table}

For the $B_s \to VV$ decays, we also give the polarization fractions
in addition to the branching ratios and CP asymmetry
parameters.\cite{PQCDBs} Similar to the $B\to VV$ decay channels, we
also have large transverse polarization fractions for the penguin
dominant processes, such as $B_s \to \phi\phi$, $B_s\to
K^{*+}K^{*-}$, $K^{*0}\bar K^{*0}$ decays, whose transverse
polarization fraction can reach 40-50\%.

\section{SU(3) breaking effect}

The SU(3) breaking effects comes mainly from the $B_s(B_d)$ meson
decay constant and distribution amplitude parameter, light meson
decay constant and wave function difference, and various decay
topology differences. As an example we mainly focus on  the decays
$B \to \pi \pi$, $B \to K \pi$, $B_s \to K\pi$ and $B_s \to KK$, as
they can be related by SU(3)-symmetry. A question of considerable
interest is the amount of SU(3)-breaking in various topologies
(diagrams) contributing to these decays. For this purpose, we
present in Table~\ref{Amplitudes} the magnitude of the decay
amplitudes (squared, in units of GeV$^2$) involving the distinct
topologies for the four decays modes. The first two decays in  this
table are related by U-spin symmetry $(d \to s)$ (likewise the two
decays in the lower half). We note that the assumption of U-spin
symmetry for the (dominant) tree ($\mathcal{T}$) and penguin
($\mathcal{P}$)
 amplitudes in
the emission diagrams is quite good, it is less so in the other
topologies, including the contributions from the $W$-exchange
diagrams, denoted by $\mathcal{E}$ for which there are non-zero
contributions for the flavor-diagonal states $\pi^+\pi^-$ and
$K^+K^-$ only. The U-spin breakings are large
 in the electroweak penguin induced amplitudes $\mathcal{P_{EW}}$,
and in the penguin annihilation amplitudes $\mathcal{P_A}$ relating
the decays $B_d \to K^+\pi^-$ and $B_s \to K^+ K^-$.
 In the SM, however,
the amplitudes $\mathcal{P_{EW}}$ are negligibly small.

\begin{table}[tbh]
\caption{Contributions from the various topologies to the decay
amplitudes (squared) for the four indicated decays. Here,
$\mathcal{T}$ is the contribution from the color favored emission
diagrams; $\mathcal{P}$ is the penguin contribution from the
emission diagrams; $\mathcal{E}$ is the contribution from the
W-exchange diagrams; $\mathcal{P_A}$ is the contribution from the
penguin annihilation amplitudes; and $\mathcal{P_{EW}}$ is the
contribution from the electro-weak penguin induced amplitude.}

\begin{center}
 \begin{tabular}{c|ccccc}
  \hline\hline
 mode ($\mbox{GeV}^2$)& $|\mathcal{T}|^2$&$|\mathcal{P}|^2$&$ |\mathcal{E}|^2$&$|\mathcal{P_A}|^2$&$|\mathcal{P_{EW}}|^2$\\ \hline
   $B_d \to \pi^+\pi^-$   &~~~$1.5$~~~&  $9.2\times 10^{-3}$  &$6.4\times 10^{-3}$ &$7.5\times 10^{-3}$  & $2.7\times 10^{-6}$ \\

   $B_s \to \pi^+ K^- $   &~~~$1.4$~~~&  $7.4\times 10^{-3}$  & $0$                &$7.0\times 10^{-3}$  & $5.4\times 10^{-6}$
   \\ \hline\hline
   $B_d \to K^+\pi^-  $   &~~~$2.2$~~~&  $18.8\times 10^{-3}$   &  0                 &$4.7\times 10^{-3}$  & $7.4\times 10^{-6}$  \\
   $B_s \to K^+K^-    $   &~~~$2.0$~~~&  $14.7\times 10^{-3}$  &
$4.6\times 10^{-3}$& $9.8\times 10^{-3}$ & $3.1\times 10^{-6}$  \\
 \hline\hline\end{tabular}\label{Amplitudes}
\end{center}
 \end{table}

 In $\overline{B_d^0}\to K^-\pi^+$ and $\overline{B_s^0}\to
K^+\pi^-$, the branching ratios are very different from each other
due to the differing strong and weak phases entering in the tree and
penguin amplitudes. However, as shown by
Gronau~\cite{Gronau:2000zy}, the two relevant products of the CKM
matrix elements entering in the expressions for the direct CP
asymmetries in these decays are equal, and, as stressed by
Lipkin~\cite{KpiLipkin} subsequently, the final states in these
decays are charge conjugates, and the strong interactions being
charge-conjugation invariant, the direct CP asymmetry in
$\overline{B_s^0}\to K^-\pi^+$ can be related to the well-measured
CP asymmetry in the decay $\overline{B_d^0}\to K^+\pi^-$ using
U-spin symmetry.

Following the suggestions in the literature, we can  define the
following two parameters:
\begin{eqnarray}
R_3&\equiv&\frac{|A(B_s\to\pi^+K^-)|^2-|A(\bar
B_s\to\pi^-K^+)|^2}{|A(B_d\to\pi^-K^+)|^2-|A(\bar
B_d\to\pi^+K^-)|^2},\\
\Delta&=&\frac{A^{dir}_{CP}(\bar B_d\to\pi^+K^-)}{A^{dir}_{CP}(\bar
B_s\to\pi^-K^+)}+\frac{BR(B_s\to\pi^+K^-)}{BR(\bar
B_d\to\pi^+K^-)}\cdot\frac{\tau(B_d)}{\tau(B_s)}.
\end{eqnarray}
The standard model predicts $R_3=-1$ and $\Delta=0$ if we assume
$U$-spin symmetry. Since we have a detailed dynamical theory to
study the SU(3) (and U-spin) symmetry violation, we can check in
pQCD approach how good quantitatively this symmetry is in the ratios
$R_3$ and $\Delta$. We
  get $R_3=-0.96 ^{+0.11}_{-0.09}$ and $\Delta =
-0.03\pm 0.08$. Thus, we find that these quantities are quite
reliably calculable, as anticipated on theoretical grounds. SU(3)
breaking and theoretical uncertainties are very small here, because
most of the breaking effects and uncertainties are canceled due to
the definition of $R_3$ and $\Delta$. On the experimental side, the
results for $R_3$ and $\Delta$ are:~\cite{CDFBsKKrecent}
\begin{equation}
R_3=-0.84\pm0.42\pm0.15,\;\; \Delta=0.04\pm0.11\pm0.08.
\end{equation}
We conclude that SM is in  good agreement with the data, as can also
be seen in Fig.~\ref{R3} where we plot theoretical predictions for
$R_3$ vs.~$\Delta$ and compare them with the current measurements of
the same. The  measurements of these quantities are rather imprecise
at present,
 a situation which we hope will greatly improve at the LHCb experiment.

\begin{figure}
 \centerline{
\psfig{file=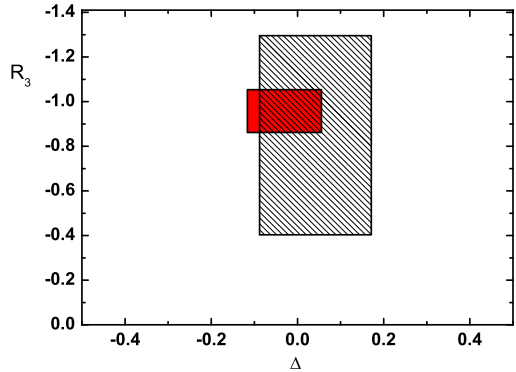,width=6.7cm,angle=0}} \caption{ $R_3$ vs
$\Delta$: The red (smaller) rectangle is the pQCD estimates worked
out in this paper. The experimental results with their $\pm 1
\sigma$ errors are shown as the larger
 rectangle. }\label{R3}
 \end{figure}

\section{Summary}

 Based on
the $k_T$ factorization, pQCD approach is infrared safe. Its
predictions on the branching ratios and CP asymmetries of the
$B^0(B^\pm)$ decays are tested well by the B factory experiments.
Using those tested parameters from these decays,  we calculate a
number of charmless decay channels $B_s\to PP$, $PV$ and $VV$ in the
perturbative QCD approach. The  experimental measurements of the
three $B_s$ decay channels are consistent with our numerical
results. Especially the measured direct CP asymmetry of $B_s\to
\pi^-K^+$ agree with our calculations. We also discuss the SU(3)
breaking effect in these decays, which is at least around 20-30\%.
We also show that the Gronau-Lipkin sum rule works quite well in the
standard model, where the SU(3) breaking effects mainly cancel.

\section*{Acknowledgments}
We are grateful to the collaborators of this work: A. Ali, G.
Kramer, Y. Li, Y.L. Shen, W. Wang and Y.M. Wang. This work is partly
supported by National Science Foundation of China under Grant
No.~10475085 and 10625525.

\end{document}